\documentclass[aps,twocolumn,pre,superscriptaddress,showpacs]{revtex4-1}

\usepackage{graphicx}
\usepackage{dcolumn}
\usepackage{bm}

\usepackage{amstext,amsmath}

\begin{document}


\title{New model of calculating the energy transfer efficiency \\for the spherical theta-pinch device}

\author{G. Xu}\altaffiliation{xuge@impcas.ac.cn}
\affiliation
{
Institute of Modern Physics, Chinese Academy of Sciences, Lanzhou,
730000, China
}
\affiliation
{
University of Chinese Academy of Sciences, Beijing, 100049, China
}
\affiliation
{
Plasma physics Group, Institute of Applied Physics, Goethe University, 60438,
Frankfurt am Main, Germany
}

\author{C. Hock}
\affiliation
{
Plasma physics Group, Institute of Applied Physics, Goethe University, 60438,
Frankfurt am Main, Germany
}
\author{G. Loisch}
\affiliation
{
Plasma physics Group, Institute of Applied Physics, Goethe University, 60438,
Frankfurt am Main, Germany
}
\author{G. Xiao}
\affiliation
{
Institute of Modern Physics, Chinese Academy of Sciences, Lanzhou,
730000, China
}
\author{J. Jacoby}
\affiliation
{
Plasma physics Group, Institute of Applied Physics, Goethe University, 60438,
Frankfurt am Main, Germany
}
\author{K. Weyrich}
\affiliation
{Plasmaphysik, GSI-Darmstadt, 64291, Darmstadt, Germany
}
\author{Y. Li}
\affiliation
{
Institut f\"ur Theoretische Physik, Goethe-Universit\"at Frankfurt am Main,
60438 Frankfurt am Main, Germany
}

\author{Y. Zhao}
\affiliation
{
Institute of Modern Physics, Chinese Academy of Sciences, Lanzhou,
730000, China
}

\date{\today}

\begin{abstract}
Ion-beam-plasma-interaction plays an important role in the field of Warm Dense Matter (WDM) and Inertial Confinement Fusion (ICF). A spherical theta pinch is proposed to act as a plasma target in various applications including a plasma stripper cell. One key parameter for such applications is the free electron density. A linear dependency of this density to the amount of energy transferred into the plasma from an energy storage was found by C. Teske. Since the amount of stored energy is known, the energy transfer efficiency is a reliable parameter for the design of a spherical theta pinch device. The traditional two models of energy transfer efficiency are based on assumptions which comprise the risk of systematical errors. To obtain precise results, this paper proposes a new model without the necessity of any assumption to calculate the energy transfer efficiency for an inductively coupled plasma device. Further, a comparison of these three different models is given at a fixed operation voltage for the full range of working gas pressures. Due to the inappropriate assumptions included in the traditional models, one  owns a tendency to overestimate the energy transfer efficiency whereas the other leads to an underestimation. Applying our new model to a wide spread set of operation voltages and gas pressures, an overall picture of the energy transfer efficiency results.
\end{abstract}

\maketitle

\section{\label{sec:level1}Introduction}
In former works Z-pinches have been used as plasma targets in scientific fields like the research on Inertial Confinement Fusion, High Energy Density Physics and accelerator development \cite{Hoffmann1990,Dietrich1992,Jacoby1995,Weyrich1989,Hoffmann1994,YongtaoZhao2012}. Plenty of achievements have been obtained in these scientific fields. As the Z-pinch is a capacitive discharge between two electrodes, erosion plays an important role in the lifetime limitation of such devices. Therefore the Z-pinch is not the optimal choice for high repetition rate applications e.g. as a plasma stripper. However, an inductively coupled plasma device has the advantage to use external coils without contact with the plasma to initiate and maintain the gas ionization. The so called theta-pinch combines the high lifetime  with the free electron density comparable to those of Z-pinches \cite{Teske2008,Teske2010,Teske2012}.

Conventional small scale theta pinch devices have been broadly investigated in various applications \cite{Aisenberg1964,Luna1998}. The maximum energy transfer efficiency was determined to be 59\% \cite{Silberg1966}. However, in 2008, a spherical theta pinch with the significantly advanced  energy transfer efficiency up to 80\% was realized by C. Teske and J. Jacoby \cite{Teske2008}. Later on, the efficiency was raised to 86\% \cite{Teske2012}. All these results for the energy transfer efficiency were calculated out by applying the model first proposed by S. Aisenberg in 1964 \cite{Aisenberg1964}.

The key parameter for the plasma stripper is the free electron density. To achieve the higher free electron density, the method is to upgrade the device. However, the remaining difficulty is how to forecast the electron density after upgrading. Fortunately, a linear scaling law between  the free electron density and the deposited energy in plasma was shown in \cite{Teske2012}. To derive the electron density, the knowledge of the precise energy transfer efficiency values is essential. Reviewing the former energy transfer efficiency models \cite{Aisenberg1964,Silberg1966,Luna1998,Cavalcanti2009,Teske2010}, all of them contains an inaccurate assumption that the reflected plasma resistant is  constant. Such treatment is to solve the differential equation which will be well described in this article.   To obtain the precise results, a new model based on the experimental measurement is proposed.

\section{\label{sec:level2}Experimental setup}

\begin{figure}
\includegraphics[width=0.45\textwidth]{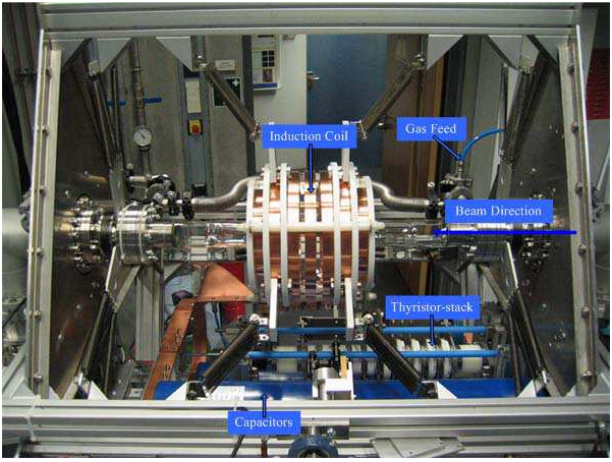}
\caption{\label{device}Spherical theta-pinch setup for the plasma stripper application}
\end{figure}

The basic working principle and experimental setup have been described already in preceding papers \cite{Teske2008,Teske2010,Teske2012,Teske2009,Teske2010IEEE}. In this article, to apply as a plasma stripper, the spherical theta pinch is    reconstructed and upgraded as shown in fig.~\ref{device}. The device axis is positioned horizontally to match the ion beam direction. To achieve a higher electron density, six capacitors with each a capacitance of 25 \(\mu\)F are used, being connected parallelly in three groups of two capacitors connected in series. This setting results in a total capacitance of 37.5 \(\mu\) F with a maximum charging voltage of 18 kV. To satisfy the demand of a long life-time even at high repetition rates, the maximum operation voltage is limited to 14 kV. Nevertheless, this means a maximum of 3.7 kJ stored energy in the capacitors which is almost twice as much as in the previous setup \cite{Teske2012} . Correspondingly,  the thyristor-stack is also improved for adapting the operation voltage of 14 kV.  Besides, the bigger glass vessel of the 6000 \(cm^{-3}\) discharge volume is closely encircled by a seven-turns coil. The Capacitors, the switch, the transmission line and the coil form a  typical LRC (inductance, resistance and capacitance) circuit with a resonance frequency of about 10 kHz.

A mixture of mainly Argon and 2.2\% Hydrogen is used as a working gas and the pressure is measured by a full range pressure gauge. As a diagnostic tool, a fast photo diode is installed to monitor the plasma. In addition, a Rogowski coil measures the circuit current in the transmission line. These two diagnostic signals are recorded by an oscilloscope setting at two sample rates of 1 and 2.5 MHz.

\section{\label{sec:level3}Models of energy transfer efficiency}

\begin{figure}
\includegraphics[width=0.3\textwidth]{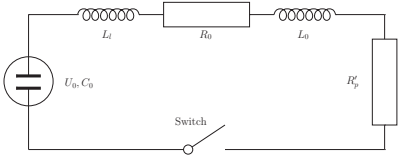}
\caption{\label{fig:circuit}LRC equivalent circuit for the plasma stripper set-up}
\end{figure}

\begin{figure}
\includegraphics[width=0.45\textwidth]{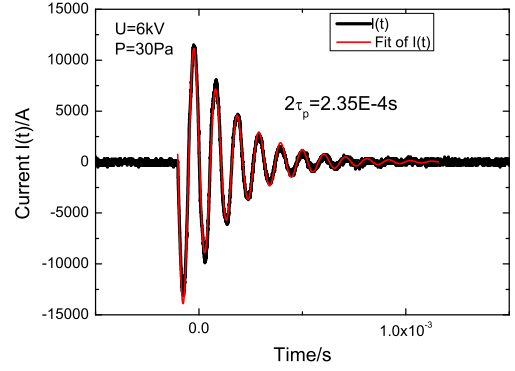}
\caption{\label{fig:typical_current}Typical current signal shape and corresponding fit}
\end{figure}

\begin{figure}
\includegraphics[width=0.45\textwidth]{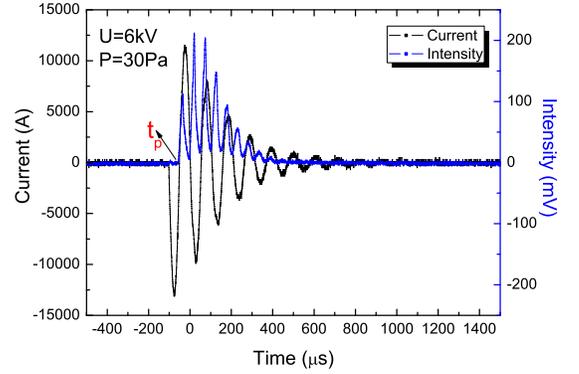}
\caption{\label{fig:light_signal}Light signal from the plasma with assumed plasma start time}
\end{figure}

\begin{figure}
\includegraphics[width=0.45\textwidth]{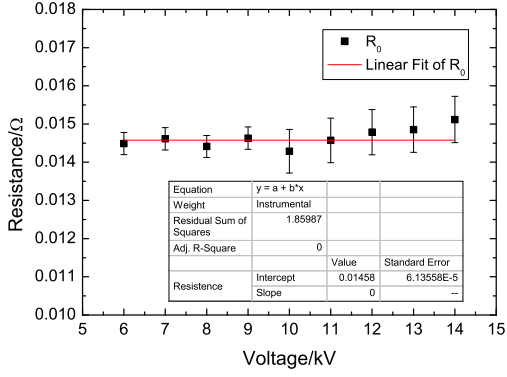}
\caption{\label{fig:parasitic_resistance}Experimental value for the parasitic resistance}
\end{figure}

To calculate the energy transfer efficiency of theta pinch devices, S. Aisenberg et al. first proposed an equivalent circuit as Fig.~\ref{fig:circuit} \cite{Aisenberg1964,Silberg1966}. This is a typical LRC circuit. In this circuit, $L_l$ and $R_0$ are undesired parasitic inductances and resistances respectively. $L_0$ is the inductance of the coil while $R_p^{'}$ represents the reflected plasma resistance which is the transformed value from the real plasma resistance into the LRC circuit. $C_0$ is the total capacitance of the capacitor bank with the initial charging voltage $U_0$. Under the frame of this equivalent circuit, the traditional and the new models are derived as follows.

\subsection{\label{sec:level31}Traditional models}
\subsubsection{\label{sec:level311}S. Aisenberg's model}
As Fig.~\ref{fig:circuit}  fulfils Kirchhoff's voltage law, the circuit equation is expressed as
\begin{eqnarray}
(\ L_{0}+L_l)\frac{dI(t)}{dt}+(R_0 +R_{p}^{'})I(t)+\frac{\int^{t}_0 I(t')dt'}{C_0}=0
\label{eq:circuit}.
\end{eqnarray}
To solve this differential equation, $R_{p}^{'}$ is assumed to be constant during the whole discharge. Then Eq.~\eqref{eq:circuit} is transformed to
\begin{eqnarray}
(\ L_{0}+L_l)\ddot I(t)+(R_0+R_p^{'})\dot I(t)+\frac{1}{C_0}I(t)=0
\label{eq:derivedcircuit},
\end{eqnarray}
Generally, there are three different solutions depending on the value of $\Delta=(R_0+R_p^{'})^{2}-4\frac{L_{0}+L_l}{C_0}$:
\begin{eqnarray}
I(t)&=&I_0e^{-\frac{R_0+R_p'}{2(L_0+L_l)}t}\sin\left(\frac{\sqrt{-\Delta}}{2(L_0+L_l)}t\right),  ~\Delta<0,
\label{eq:solve1}\\
I(t)&=&I_1te^{-\frac{R_0+R_p'}{2(L_0+L_l)}t},~\Delta=0,
\label{eq:solve2}\\
I(t)&=&I_2e^{-\frac{R_0+R_p'}{2(L_0+L_l)}t}\left[e^{\frac{\sqrt{\Delta}}{2(L_0+L_l)}t}-e^{{\frac{-\sqrt{\Delta}}{2(L_0+L_l)}t}}\right],~\Delta>0. ~~~~
\label{eq:solve3}
\end{eqnarray}
Considering the current curve shape  in Fig.~\ref{fig:typical_current},  which is a damped oscillation, Eq.~\eqref{eq:solve1} is the only proper solution which is simplified to
\begin{eqnarray}
I(t)=I_0e^{-\frac{t}{2\tau_p}}\sin(\omega t)
\label{eq:simplify solution}
\end{eqnarray}
with
$\tau_p={\frac{L_0+L_l}{R_0+R_p'}},~\omega={\frac{\sqrt{-\Delta}}{2(L_0+L_l)}},~I_0=\frac{U_0}{\omega(L_0+L_l)}$,
where $2\tau_p$ is the decay constant while $\omega$ is the oscillation frequency of the circuit. In the case of $(R_0+R_p')^2\ll4\frac{L_0+L_l}{C_0}$, $\omega$ is approximated as
\begin{eqnarray}
\omega=\sqrt{\frac{1}{C_0(L_0+L_l)}}.
\end{eqnarray}

Further, the energy consumed in the plasma is
\begin{eqnarray}
W_p&=&\int^{+\infty}_{0}I^2(t){R_p'}dt\nonumber\\
&=&\frac{1}{2}I_0^2{R_p'}\tau_p\left(1-\frac{1}{4\omega^2\tau_p^2+1}\right).
\label{eq:model1_plasma_energy}
\end{eqnarray}
In the case of $\omega \tau_p>5$,
the approximation for $W_p$ within an error of 1\% is
\begin{eqnarray}
W_p\approx\frac{1}{2}I_0^2{R_p'}\tau_p.
\end{eqnarray}

Analogously, for a free oscillation, the energy dissipated in the parasitic resistance equals the whole stored energy $E$:
\begin{equation}
\begin{split}
E=\int^{+\infty}_{0}I^2(t){R_0}dt\approx\frac{1}{2}I_0^2{R_0}\tau_0,
\end{split}
\end{equation}
where $2\tau_0$ is the decay constant for the free oscillation.

Finally, the energy transfer efficiency $\eta$ from the capacitors to the plasma is
\begin{eqnarray}
\eta=\frac{W_p}{E}=1-\frac{\tau_p}{\tau_0}.
\label{eq:model1}
\end{eqnarray}

\subsubsection{\label{sec:level312}C.Teske's modification}
In this modification, it is assumed that the plasma start time $t_p$  indicated by the light emission from the plasma is not the same as the current. Hence, the energy transferred to the plasma is
\begin{eqnarray}
W_p&=&\int^{+\infty}_{t_p}I^2(t){R_p'}dt\nonumber\\
&=&\frac{1}{2}I_0^2{R_p'}\tau_p\exp\left(-\frac{t_p}{\tau_p}\right)\bigg[1+\frac{2\omega\tau_p}{4\omega^2\tau_p^2+1}\sin(2\omega t_p)\nonumber\\
&&-\frac{1}{4\omega^2\tau_p^2+1}\cos(2\omega t_p)\bigg].
\label{eq:model2_plasma_energy}
\end{eqnarray}
Usually, the plasma ignites after the first half wave which suggests $t_p\approx\pi/\omega$ as Fig.~\ref{fig:light_signal} shows. Besides, $\omega \tau_p$ is assumed to be much greater than 1. As a result, the energy deposited in the plasma is approximated as
\begin{eqnarray}
W_p\approx\frac{1}{2}I_0^2{R_p'}\tau_p\exp\left(-\frac{t_p}{\tau_p}\right)
\end{eqnarray}
Correspondingly, the modified energy transfer efficiency formula is attained
as \begin{eqnarray}
\eta=\left(1-\frac{\tau_p}{\tau_0}\right)\exp\left(-\frac{t_p}{\tau_p}\right).
\label{eq:model2}
\end{eqnarray}

\subsection{\label{sec:level32}New model}
In the preceding models, the expressions for the energy transfer efficiency are both based on the assumption that the reflected plasma resistance is constant. Since the light signal oscillates with time shown in Fig.~\ref{fig:light_signal}, neither the plasma parameters nor the value of the reflected plasma resistance can be constant. Hence, the traditional models contain the risk of deviating from the true value.

To avoid this risk, a new model without any assumption is established. The key point for this new model is that the parasitic resistance is constant as Fig.~\ref{fig:parasitic_resistance} shows, for the experimentally measured values. Moreover, since the capacitance and inductance do not consume the energy, the total energy is completely dissipated in the parasitic resistance and the reflected plasma resistance. As a result, the energy deposited in the plasma is
\begin{eqnarray}
W_p=E-R_0\Delta t\sum_{i=1}^{\infty}I^2(t_i),
\end{eqnarray}
where $E$ is the total energy stored in the capacitors; $\Delta t$ is a constant time interval between $t_i$ and $t_{i+1}$; $I(t_i)$ is the measured current value at $t_i$. Even though $\Delta t$ is considered to be infinitesimally small, the sum is used to take into account the finite time resolution of the oscilloscope. For our experiment, the substitution error is less than 0.01\%.

For the free oscillation, the total energy which is completely dissipated in the parasitic resistance
is \begin{eqnarray}
E=R_0\Delta t_0\sum_{i=1}^{\infty}I_0^2(t_i).
\end{eqnarray}
Here, $I_0(t_i)$ is the measured current value at $t_i$. $\Delta t_0$ is again a constant time interval between $t_i$ and $t_{i+1}$.

Thus, the energy transfer efficiency is written as
\begin{eqnarray}
\eta=1-\frac{\Delta t\sum_iI^2(t_i)}{\Delta t_0\sum_iI_0^2(t_i)}.
\label{eq:model3}
\end{eqnarray}

\section{\label{sec:level4}Results and discussion}

\begin{figure}
\includegraphics[width=0.45\textwidth]{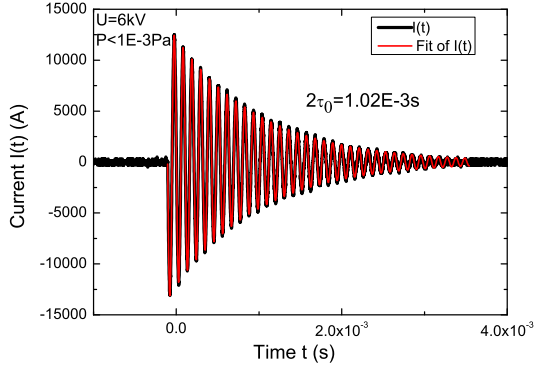}
\caption{\label{Decay_constant}Decay constant for the case of free oscillation at 6 kV}
\end{figure}

\begin{figure}
\includegraphics[width=0.45\textwidth]{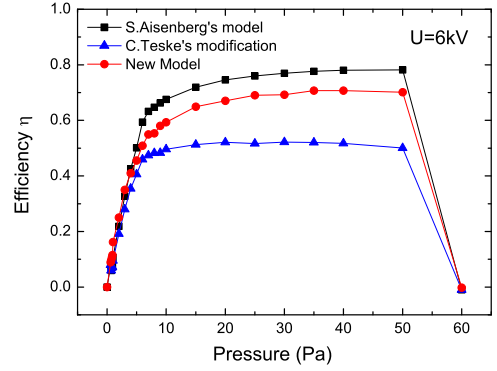}
\caption{\label{Three_models}Efficiencies obtained with the three methods at 6 kV}
\end{figure}

\begin{figure}
\includegraphics[width=0.45\textwidth]{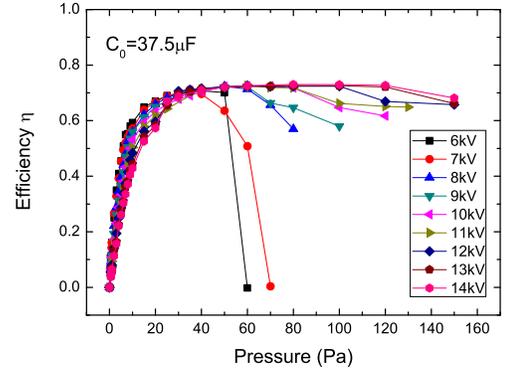}
\caption{\label{Energy_transfer_efficiencies}Energy transfer efficiencies of a complete pressure and voltage set}
\end{figure}

In this article, an operation voltage is applied from 6 to 14 kV. For each operation voltage, the gas pressure varies in the range of 0.6-160 Pa and an additional "zero shot" in the order of $10^{-3}$ Pa. This "zero shot" represents the free oscillation of the current.  The black line in Fig.~\ref{Decay_constant} shows this kind of  free  oscillation at 6 kV. The red line displaying a good overlap with the black one is the fitting curve. Gained from the fitting parameters, the decay constant of this free oscillation is 1.02 ms. Analogously, the decay constants for the different pressures are obtained.

Directly applying Eq.~\eqref{eq:model1} from the S. Aisenberg's model on these decay constants, the corresponding energy transfer efficiencies for the different gas pressures are calculated. For the C. Teske's modified model, another parameter besides the decay constants is the plasma start time $t_p$. This is obtained from the difference between the beginning of the light emission in time and the current start time.  Then,  adopting  Eq.~\eqref{eq:model2} , the modified energy transfer efficiencies are also computed. Both of the preceding models need the plotting and fitting procedures which cause propagating errors.
However,
for our new model, there is no need  to plot and fit the data because Eq.~\eqref{eq:model3} is straightforwardly applied on the measured current values from the data sheet. Consequently, the energy transfer efficiencies are derived from this.

A comparison is made among  the results from these models at 6 kV shown in Fig.~\ref{Three_models}. Their tendencies  are extremely similar. At low pressures, the curves rise very fast.
Then they slow down at medium pressures. Finally, all of them suddenly drop down to the point of 60 Pa which is approximated the breakdown threshold. The biggest difference among the models can be found  at medium pressures over a wide range. The S. Aisenberg model
shows the highest efficiencies whereas the C. Teske's gives the lowest in this pressure range. The maximum energy transfer efficiency for the S. Aisenberg's model is 78\% while 52\% for C. Teske's modified model. In our new model, the value for the maximum energy transfer efficiency is 71\%. Due to no assumption made in the new model, it is considered reliable and precise. Hence, the S. Aisenberg's model overestimates the energy transfer efficiency. On the contrary, the C. Teske's modification model underestimates the values.

These deviations for the traditional models are mainly caused by the assumption that the reflected plasma resistance is constant, which is untrue. The truth is the reflected plasma resistance varies with time as the light signal in Fig.~\ref{fig:light_signal} shows. However, this assumed constant reflected plasma resistance leads to an approximately average value for the true time-dependent resistance. Regarding the energy consumed in the plasma, it does not only depend on the resistance but also the square of the current. Namely, the square of the current is the weighting factor of the resistance for calculating the transferred energy. Due to the oscillation decay behavior of the current, the weighting factor for the first half wave is bigger than for the other half waves. Unfortunately, the resistance for the first half wave is almost zero indicated by the light signal shown in Fig.~\ref{fig:light_signal}. Consequently, the calculated energy transfer by applying the average resistance is overestimated. This accounts for the S. Aisenberg's model's overestimation. For C.Teske's modified model, this first half wave of the current is omitted for calculating the transferred energy according to the Eq.~\eqref{eq:model2_plasma_energy}. However, the calculation of the average resistance still counts in this first half wave where the reflected plasma resistance is almost 0. This results in the underestimation of the average resistance. Correspondingly, the transferred energy calculation is underestimated. Besides,
the neglection of the sine  term in Eq.~\eqref{eq:model2_plasma_energy} also contributes to this energy underestimation. For instance at 30 Pa and 6 kV, the maximum deviation for this sine term is about 3\%.

Due to the advantage of our new model and the faults of the traditional models, our new model is adapted to calculate the energy transfer efficiency from the measured current values. Fig.~\ref{Energy_transfer_efficiencies} shows all the measured energy transfer efficiencies vary with the pressures of the different operation voltages. In our measured pressure range, the breakdown threshold is only observed at the lower voltage of 6 and 7 kV. For each voltage, the pressure corresponding to the peak value of the efficiency is defined as the optimal pressure which shows the maximum transfer efficiency. Comparing all of the plots in Fig.~\ref{Energy_transfer_efficiencies} , it is found that the optimal gas pressure shifts to higher values when the voltage is increased. The maximum  energy transfer efficiency found here is 73\% which occurs at around 80 Pa under the maximum voltage of 14 kV.\\

\section{\label{sec:level5}Conclusion}

A new model is proposed to calculate the energy transfer efficiency of theta pinch in general and especially for the spherical theta pinch. As this new model contains no assumptions, the results obtained with it are considered exact.

The deviations of the traditional models of S.Aisenberg and C.Teske from the real values are mainly caused by  the improper assumption that the reflected plasma resistance is constant.
The first leads to overestimation whereas the latter to underestimation. The corresponding  explanations are given respectively.

The latest setup of the spherical theta pinch that is used for the presented investigations has a maximum energy transfer efficiency of 73\% for an Argon-Hydrogen gas mixture which is promising especially for high repetition rate applications e.g. flash lamps or a plasma stripping device.

\begin{acknowledgments}
This work is supported by the BMBF (German Ministry for Education and Science) under the contract number 05P12RFRB8. The authors Ge Xu and Christion Hock are supported by scholarships from HGS-HIRe (Helmholtz Graduate School for Hadron and Ion Research) for FAIR (Facility for Antiproton and Ion Research). We especially acknowledge Dr. Christian Teske and his colleagues for designing and manufacturing the spherical theta pinch device.
We sincerely thank our colleagues who gave us support in this work.\end{acknowledgments}


\begin{thebibliography}{10}

\bibitem{Hoffmann1990}
D.~H.~H. Hoffmann {\em et~al.},
\newblock Phys.  Rev.  A {\bf 42}, 2313 (1990).

\bibitem{Dietrich1992}
K.-G. Dietrich {\em et~al.},
\newblock Phys.  Rev.  Lett.  {\bf 69}, 3623 (1992).

\bibitem{Jacoby1995}
J.~Jacoby {\em et~al.},
\newblock Phys.  Rev.  Lett.  {\bf 74}, 1550 (1995).

\bibitem{Weyrich1989}
K.~Weyrich {\em et~al.},
\newblock Nuclear Instruments and Methods in Physics Research Section A:
  Accelerators, Spectrometers, Detectors and Associated Equipment {\bf 278}, 52
   (1989).

\bibitem{Hoffmann1994}
D.~Hoffmann {\em et~al.},
\newblock Nuclear Instruments and Methods in Physics Research Section B: Beam
  Interactions with Materials and Atoms {\bf 90}, 1  (1994).

\bibitem{YongtaoZhao2012}
Y.~Zhao {\em et~al.},
\newblock Laser and Particle Beams {\bf 30}, 679 (2012).

\bibitem{Teske2008}
C.~Teske and J.~Jacoby,
\newblock Plasma Science, IEEE Transactions on {\bf 36}, 1930 (2008).

\bibitem{Teske2010}
C.~Teske, J.~Jacoby, F.~Senzel, and W.~Schweizer,
\newblock Physics of Plasmas (1994-present) {\bf 17}, 043501 (2010).

\bibitem{Teske2012}
C.~Teske, Y.~Liu, S.~Blaes, and J.~Jacoby,
\newblock Physics of Plasmas {\bf 19}, 033505 (2012).

\bibitem{Aisenberg1964}
D.~V.~M. S.~Aisenberg and P.~A. Silberg,
\newblock J.  Appl.  Phys.  {\bf 35}, 3625 (1964).

\bibitem{Luna1998}
F.~R.~T. Luna, G.~H. Cavalcanti, and A.~G. Trigueiros,
\newblock Journal of Physics D: Applied Physics {\bf 31}, 866 (1998).

\bibitem{Silberg1966}
P.~Silberg,
\newblock Journal of Applied Physics {\bf 37}, 2155 (1966).

\bibitem{Cavalcanti2009}
G.~H. Cavalcanti and E.~E. Farias,
\newblock Review of Scientific Instruments {\bf 80}, 125109 (2009).

\bibitem{Teske2009}
C.~Teske, J.~Jacoby, W.~Schweizer, and J.~Wiechula,
\newblock Review of Scientific Instruments {\bf 80}, 034702 (2009).

\bibitem{Teske2010IEEE}
C.~Teske, B.-J. Lee, A.~Fedjuschenko, J.~Jacoby, and W.~Schweizer,
\newblock Plasma Science, IEEE Transactions on {\bf 38}, 1675 (2010).

\end{thebibliography}

\end{document}